# Silicon waveguide modulator with embedded phase change material

**KEVIN J. MILLER,[1] KENT A. HALLMAN,[2] RICHARD F. HAGLUND,[1,2] AND SHARON M. WEISS[1,2,3]**

[1]*Interdisciplinary Graduate Program in Materials Science, Vanderbilt University, Nashville, Tennessee 37235, USA*
[2]*Department of Physics and Astronomy, Vanderbilt University, Nashville, Tennessee 37235, USA*
[3]*Department of Electrical Engineering and Computer Science, Vanderbilt University, Nashville, Tennessee 37235, USA*
*\*sharon.weiss@vanderbilt.edu*

**Abstract:** Phase-change materials (PCMs) have emerged as promising active elements in silicon (Si) photonic systems. In this work, we design, fabricate, and characterize a hybrid Si-PCM optical modulator. By integrating vanadium dioxide (a PCM) within a Si photonic waveguide, in a non-resonant geometry, we demonstrate ~ 10 dB broadband modulation with a PCM length of 500 nm.

## 1. Introduction

In the Information Age, increasingly complex data streams require the continual development of smaller, faster, and more efficient information processing systems. Over the past few decades, long distance communication has transitioned from electronics-based to optics-based systems (e.g., transatlantic optical fibers) to meet these requirements. Currently, optical technologies are being realized over shorter distances (e.g., within data centers). To achieve optical information processing over even shorter distances, there is ongoing research to develop and implement light sources, modulators, and detectors for interchip and intrachip applications. Silicon (Si) photonics has emerged as a promising materials platform for interchip and intrachip communication primarily due to the existing infrastructure and knowledge base for Si processing residing in the microelectronics industry. Due to the modest tunability of Si's optical properties, realizing Si photonic modulators has either required the use of resonant structures (e.g., ring resonators and photonic crystal nanocavities) [1, 2] or large footprints in non-resonant Mach-Zehnder interferometer configurations [3]. Hybrid Si photonic modulator geometries open the door to superior optical tunability while still maintaining integration compatibility with microelectronics technology by utilizing Si photonic structures as the information carrier. Promising materials for hybrid integration in Si photonic modulators include electro-optic polymers [4, 5], plasmonic materials [6], two-dimensional materials [7], germanium [8], and phase-change materials (PCMs) [9-16]. Of these materials, PCMs show the most versatile optical tunability ($\Delta n > 1$, $\Delta \kappa \sim 10$), therefore allowing for small modulator footprints without the use of a resonant structure.

In this work, we theoretically and experimentally present a new geometry to more effectively utilize the large optical tunability of a PCM to achieve an ultrasmall, large bandwidth, high modulation depth Si optical modulator. By integrating the PCM within a Si waveguide, we maximize the interaction of the guided mode in Si with the PCM, similar to previous work with low-index, electro-optic polymers in slot waveguide geometries [17]. To our knowledge, this is the first experimental demonstration of a photonic geometry that includes a PCM within a Si photonic structure. We choose to use vanadium dioxide ($VO_2$) as the PCM in this work, although $Ge_2Sb_2Te_5$ (GST) or other PCMs could also be implemented in a similar approach, as further discussed at the end of Section 6. A correlated material, $VO_2$ demonstrates a thermally induced semiconductor-metal transition (SMT) at 68°C resulting in significant changes in electrical conductivity (up to 5 orders of magnitude change in single crystals) [18] and optical properties (at 1550 nm, n = 3.3 [semiconductor], 1.8 [metal], and κ = 0.3 [semiconductor], 3.3 [metal]) [19]. This SMT can also be driven with an applied electric field [20], in which its semiconductor-metal and metal-semiconductor transitions occur on time scales less than 2 ns [14, 21] and 3 ns [14], respectively. Under femtosecond optical excitation, simultaneous time-resolved electron diffraction and optical spectroscopy have revealed the existence of a transient state in which the optical contrast of $VO_2$, while reduced in magnitude, can be induced nearly instantaneously without atomic rearrangement [22, 23]. For incident fluences of 2-3 mJ/cm$^2$, recovery times of ~ 1 and 10 ps have been demonstrated [22, 24]. For these reasons, $VO_2$ is a promising PCM for high-speed Si-PCM electro-optic and all-optical modulators.

## 2. Simulation

All simulations were performed using three-dimensional finite-difference time-domain (FDTD) analysis (Lumerical FDTD Solutions). Simulations were run with an auto non-uniform meshing parameter of 6 and perfectly matched layer boundary conditions for the fundamental TE mode source. The simulated structure consisted of a Si ridge waveguide on an $SiO_2$ substrate with a $VO_2$ patch of the same cross sectional dimensions as the Si ridge embedded within the Si waveguide, as shown in Fig. 1a. The width of the waveguide and $VO_2$ patch was 700 nm and the depth was 220 nm. The length of the $VO_2$ patch in the direction of propagation ($L_{VO_2}$) was varied from 0 to 1000 nm in steps of 100 nm. Transmission through each waveguide was calculated using frequency-domain field and power monitors; both the semiconducting and metallic states of $VO_2$ were considered. Optical constants for $VO_2$ were taken from [19] and imported into Lumerical for the simulations. Fig. 1b shows transmission as a function of $L_{VO_2}$ for both semiconducting and metallic $VO_2$ at 1550 nm, presenting individual simulation results using blue circles (semiconducting $VO_2$) and red squares (metallic $VO_2$). The curve fit for the transmission data with semiconducting $VO_2$ patches is a single exponential function obeying Beer's law for κ = 0.31, which is in good agreement with the optical properties of semiconducting $VO_2$ films at 1550 nm. However, the curve fit for the transmission data with metallic $VO_2$ patches does not simply follow Beer's law for light transmitted through an equivalent thickness $VO_2$ thin film. The spread in the optical mode profile as light passes through the low refractive index metallic $VO_2$ section of the waveguide causes a significant portion of the mode to propagate in the cladding region outside the lossy $VO_2$ patch. Insertion loss and modulation depth of the $VO_2$ embedded Si waveguide modulator are calculated from the transmission data in Fig. 1b. Insertion loss (Fig. 1c) is reported based on the transmission of the $VO_2$ embedded Si waveguide modulator when the $VO_2$ patch is in the semiconducting state relative to the transmission of a control Si ridge waveguide with no trench and no $VO_2$. Modulation depth (Fig. 1d) is reported based on the ratio of transmission through the $VO_2$ embedded Si waveguide modulator for $VO_2$ in the semiconducting and metallic states. The trend in modulation depth as a function of $VO_2$ length that shows a maximum modulation depth for a $VO_2$ patch length of 400 nm can be explained by considering the saturation in the transmission intensity of light through the $VO_2$ embedded Si waveguide modulator for metallic $VO_2$ patch lengths greater than about 400 nm. Taking into account both modulation depth and insertion loss, Fig. 1 suggests that a favorable geometry is a 200 nm embedded $VO_2$ patch, which enables nearly 14 dB modulation depth with approximately 2 dB insertion loss.

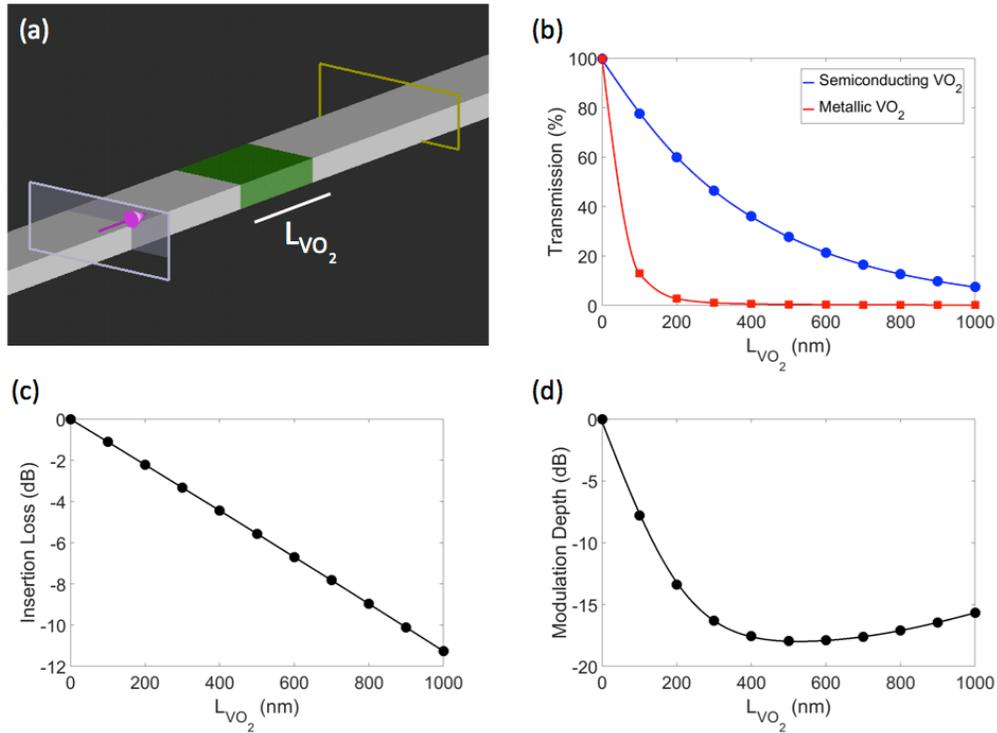

Fig. 1. (a) Schematic for simulation of $VO_2$ embedded Si waveguide modulator. The $VO_2$ embedded Si waveguide modulator is shown in gray (Si) and green ($VO_2$). The purple arrow surrounded by the gray box and yellow box represent the optical source and monitor, respectively. (b) Simulated transmission as a function of $L_{VO_2}$ through $VO_2$ embedded Si waveguide modulator with $VO_2$ in its semiconducting (blue circles) and metallic (red squares). Optical properties of $VO_2$ are taken from [19]. (c) Insertion loss and (d) modulation depth of $VO_2$ embedded Si waveguide modulator as a function of $L_{VO_2}$, calculated from transmission data shown in (b). The corresponding curve fits in (b), (c), and (d) serve as guides to the eye.

### 3. Fabrication

A scanning electron microscopy (SEM) image of the $VO_2$ embedded Si waveguide modulators is shown in Fig. 2. The input waveguides (on right side of image) are bifurcated using a 50/50 splitter such that the upper waveguide in each pair includes a trench in the waveguide for subsequent $VO_2$ deposition (Fig. 2 top left inset), while the lower waveguide is a standard ridge waveguide for control measurements (Fig. 2 top right inset). The $VO_2$ embedded Si integrated waveguide modulators were fabricated on silicon-on-insulator wafers (Soitec: 220 nm device layer, 3 μm buried oxide layer) using standard nanofabrication procedures. Silicon waveguides with and without trenches were defined by electron beam lithography (JEOL 9300FS 100kV) and subsequent reactive ion etching using a $C_4F_8/SF_6/Ar$ gas mixture (Oxford Plasmalab 100). Localized $VO_2$ deposition in the trenches (Fig. 2 – bottom left inset) was achieved by opening windows of the requisite dimensions using electron beam lithography (Raith eLine) and depositing $VO_x$ via sputtering of vanadium metal at 6 mTorr total pressure with 20 sccm Ar and 1 sccm $O_2$. After liftoff, the devices were annealed for 7 minutes at 450°C at 250 mTorr of $O_2$ to form polycrystalline $VO_2$. The $VO_2$ deposition process was performed in two identical iterations to ensure complete $O_2$ diffusion during the anneal step. Profilometry on a 100 μm × 100 μm square of $VO_2$ patterned on a separate region of the wafer revealed a cumulative $VO_2$ thickness of 250 nm. The thickness of the deposited $VO_2$ films embedded in waveguides with differing trench lengths is discussed in

Section 4.1. Resistive heaters (false colored gold in Fig. 2) were fabricated adjacent to the waveguides using electron beam lithographic patterning (Raith eLine), thermal evaporation of a 5 nm Cr adhesion layer and a 120 nm Au layer, and liftoff. Additionally, another set of bifurcated Si waveguides was fabricated with $VO_2$ deposited on top of the control waveguides in order to compare their performance with the Si waveguides having embedded $VO_2$ films.

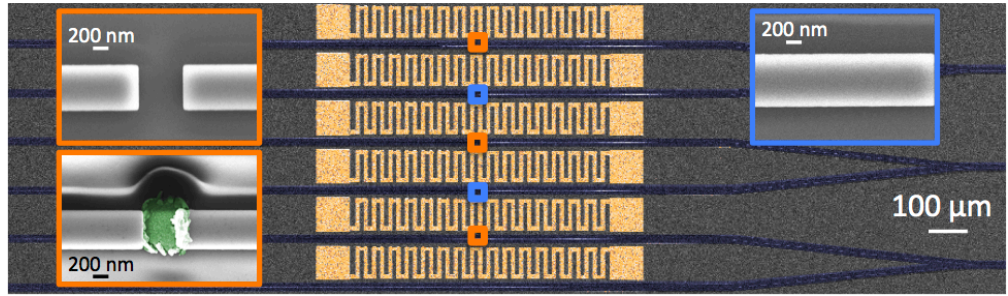

Fig. 2. SEM image of bifurcated Si waveguides (false colored navy) including integrated heaters (false colored gold). In the center of the figure, the small boxes highlight $VO_2$ embedded Si waveguide modulators (orange) and control waveguides (light blue). The left insets with orange outline show SEM images after the patterning of the Si waveguide and trench (top) and backfilling with $VO_2$ (bottom), which is shown in false colored green. The right inset outlined in light blue shows the control Si waveguide.

## 4. Characterization

### 4.1 Characterization of the $VO_2$ embedded Si Waveguide Modulators

The $VO_2$ embedded Si waveguide modulators were characterized by tilted SEM imaging and atomic force microscopy (AFM). As shown in Fig. 3a, the $VO_2$ deposition was not sufficient to completely fill the trenches. The shorter trenches with higher aspect ratio of depth to width had thinner $VO_2$ films compared to the longer trenches, most likely due to known challenges with depositing material into high aspect-ratio holes. AFM measurements for a waveguide with $L_{VO_2}$ = 1000 nm suggest that the $VO_2$ film thickness in the lowest aspect-ratio trench is approximately 180 nm (Fig. 3b), which is less than the 220 nm height of the Si waveguide. We note that the deposition of $VO_2$ on top of the Si waveguide at one end of the trench (Fig. 3a) was due to a slight misalignment of the resist window during fabrication. AFM measurements also revealed that the $VO_2$ patches on top of the control waveguides have a thickness of approximately 210 nm, which is slightly less than the thickness of the 100 μm × 100 μm square $VO_2$ film due to shadowing effects of the resist.

### 4.2 Characterization of $VO_2$

To verify the optical switching properties of $VO_2$, temperature dependent transmission measurements were performed on a thin film of $VO_2$ on a $SiO_2$ substrate that underwent identical processing steps to the $VO_2$ embedded Si waveguide modulators. A fiber coupled 3000 K tungsten-halogen light source (Spectral Products – Model ASBN-W-L) was incident on the sample. The sample temperature was controlled by a Peltier heater attached to the stage and temperature dependent transmitted power was detected using an InGaAs photodetector (Thorlabs PDA10CS). Fig. 3c shows the hysteresis curves for $VO_2$ including the onset of the SMT near 65°C.

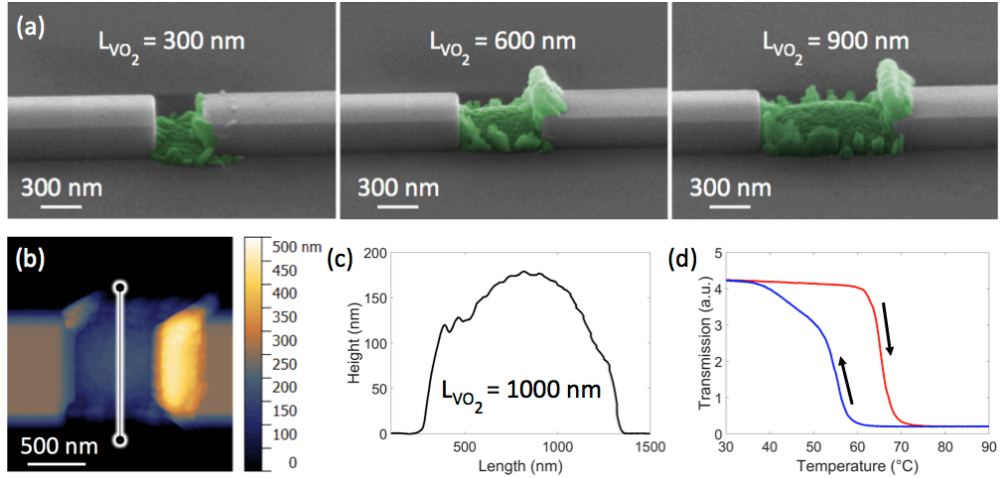

Fig. 3. (a) Tilted SEM images of VO$_2$ embedded Si waveguide modulators with $L_{VO_2}$ = 300, 600, and 900 nm. VO$_2$ is shown in false colored green. (b) AFM image of VO$_2$ embedded Si waveguide modulator with $L_{VO_2}$ = 1000 nm. Vertical profile of line cut (black line outlined in white) is presented in (c), showing a VO$_2$ thickness of ~ 180 nm within the trench. (d) Temperature dependent transmission measurements on a thin film VO$_2$ witness sample. The red and blue curves show transmission with increasing and decreasing temperature, respectively.

## 5. Transmission Measurements

Fiber-coupled transmission measurements were carried out using the Santec Swept Test System STS-510 software package with a tunable laser (Santec TSL-510) and power meter (Newport 2936-C). Near infrared light (1500-1630 nm) was coupled into and out of the waveguides using lensed tapered fibers (OZ Optics). An infrared camera (Sensors Unlimited SU320M) was used to aid alignment. Free space polarization control of the optical input was maintained using a linear polarizer and half wave plate. Transverse electric (TE) polarization was used for all measurements.

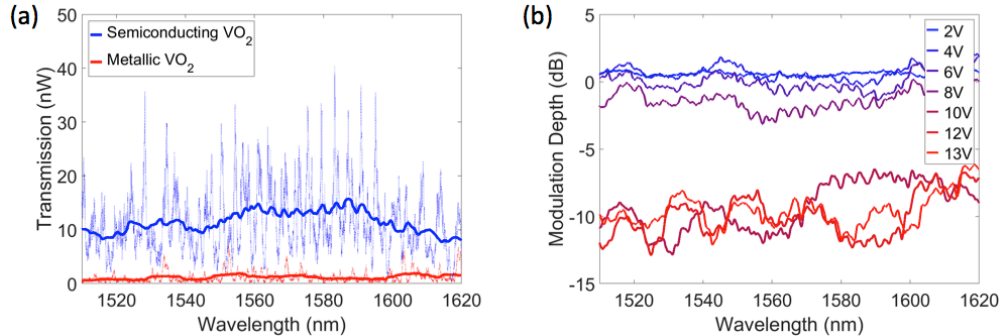

Fig. 4. (a) Raw (thin lines) and smoothed (thick lines) data for transmitted power through a VO$_2$ embedded Si waveguide modulator with $L_{VO_2}$ = 600 nm. Blue curves correspond to 0V applied (i.e., VO$_2$ in semiconducting state) and red curves correspond to 12 V applied (i.e., VO$_2$ in metallic state). (b) Modulation depth of same sample for various applied voltages, calculated from the smoothed spectra.

High contrast Fabry-Perot fringes were observed in the transmission spectra (thin lines in Fig. 4a), which are attributed to reflections from stitching errors in the waveguides that occurred during the first step of electron beam lithography, roughness at the Si-VO$_2$/Si-air interfaces, and the bifurcated waveguide design. To minimize the effect of the Fabry-Perot fringes, the data were smoothed with a moving average across 10 nm for each data point (i.e., 5 nm to each side of the data point, totaling 10,001 data points). An example of this smoothing is shown by the thick lines in Fig. 4a. These smoothed curves were used for all insertion loss and modulation depth calculations. Temperature dependent transmission measurements were carried out by varying the applied voltage across the resistive heaters. Fig. 4b shows the modulation depth, which is reported based on the ratio of transmission through the $VO_2$ embedded Si waveguide modulator with and without an applied voltage, for varying applied voltages. The SMT of $VO_2$ is initiated with approximately 8 V applied, as indicated by the change in modulation depth, and is completed with 12 V applied. Accordingly, for all calculations in the next section, we assume that $VO_2$ is in the semiconducting state when 0 V is applied to the resistive heaters and $VO_2$ is in the metallic state when 12 V is applied. Fig. 4b also demonstrates the broadband (exceeding 100 nm) operation of the $VO_2$ embedded Si waveguide modulators as expected from simulation due to the relatively small variation in optical constants of $VO_2$ from 1510-1620 nm.

## 6. Results & Discussion

After carrying out multiple transmission measurements on each of the $VO_2$ embedded Si waveguide modulators with different trench lengths, trends in the transmitted intensity, insertion loss, and modulation depth as a function of $L_{VO_2}$ were analyzed at a wavelength of 1550 nm, as shown in Fig 5. Error bars in Fig. 5 were calculated based on the standard deviation of multiple measurements on each sample. The experimental results deviate from the simulation data presented in Fig. 1b that assume complete filling of $VO_2$ in the trench because the fabricated structures have only partially filled $VO_2$ trenches (Fig. 3). Accordingly, the simulation results presented alongside the experimental data in Fig. 5 assume $VO_2$ thicknesses from 90 nm for $L_{VO_2}$ = 100 nm to 180 nm for $L_{VO_2}$ = 1000 nm, with a linear interpolation between these values (i.e., 10 nm increase in $VO_2$ height per 100 nm increase in $L_{VO_2}$). As in Fig. 1, for all simulation results shown in Fig. 5, the optical properties of $VO_2$ were taken from [19] and imported into Lumerical.

Our experimental data in Fig. 5 show good agreement with the simulated curves for Si waveguides with partially filled $VO_2$ trenches. The deviation between experiment and simulation for modulation depth (Fig. 5d) can be attributed in large part to the slightly higher measured transmission of the waveguide compared to simulation when the $VO_2$ patch is in the metallic state. The data in Fig. 5 reveal a clear tradeoff between insertion loss and modulation depth. Due to the absorption of $VO_2$ in the semiconductor state, longer $VO_2$ filled trenches lead to higher insertion losses. At the same time, the longer $VO_2$ filled trenches allow more absorption of light when $VO_2$ is in the metallic state, leading to a larger modulation depth. For the partially filled $VO_2$ embedded Si waveguide modulator with $L_{VO_2}$ = 500 nm, a modulation of 9.7 ± 0.8 dB with a corresponding insertion loss of 6.5 ± 0.9 dB is demonstrated.

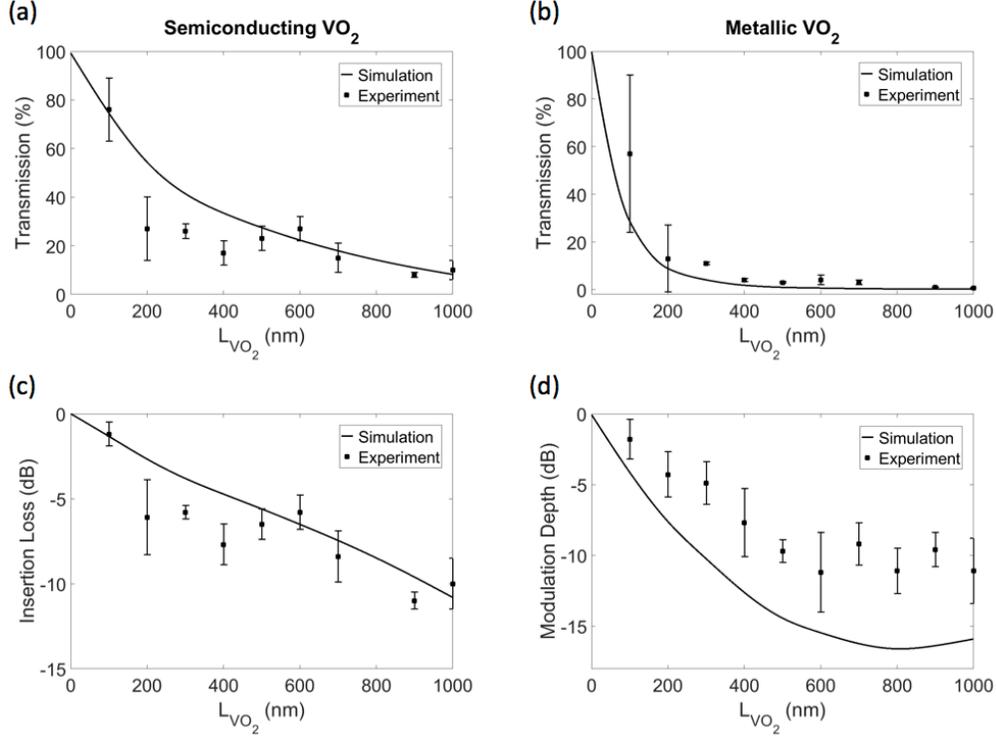

Fig. 5. Measured transmission through $VO_2$ embedded Si waveguide modulator as a function of $L_{VO_2}$, normalized to transmission through a reference Si waveguide, for $VO_2$ in its (a) semiconducting and (b) metallic state. Calculated (c) insertion loss and (d) modulation depth of $VO_2$ embedded Si waveguide modulator as a function of $L_{VO_2}$ based on measured data in (a) and (b). The solid curves present Lumerical simulation results ($VO_2$ optical properties from [19]) that assume partial $VO_2$ filling of the Si trench, as described in Section 6.

To demonstrate the advantage of placing the $VO_2$ patch within the Si waveguide instead of on top of the waveguide, we carried out the direct experimental comparison, shown in Fig. 6. As expected, the integrated geometry provides a larger modulation depth due to the improved interaction of the guided mode with the PCM. Here, the reader is reminded that the waveguides were processed in an identical manner (even though the resulting $VO_2$ thicknesses were slightly different as discussed in Section 4.1) to isolate the effect of the location of the $VO_2$ patch. However, prior work suggests that an optimized waveguide geometry may be able to further improve the modulation depth and insertion loss metrics. Using a Si-$VO_2$ rib waveguide design that supports a delocalized TE mode, 12 dB modulation with 5 dB insertion loss was reported with a 1000 nm long patch of $VO_2$ on top of the Si waveguide while 4 dB modulation was reported for a 500 nm long $VO_2$ patch [13]. We note that for our design presented in Fig. 1 which assumes complete filling of $VO_2$, our simulated results with $L_{VO_2}$ = 200 nm show modulation depth and insertion loss of 13.8 dB and 2.2 dB, respectively. Therefore, with improved Si waveguide design, in addition to improved fabrication procedures that allow more complete filling of $VO_2$ in a Si trench, it should be possible to achieve even larger modulation depths with lower insertion losses using the $VO_2$ embedded Si waveguide modulator platform.

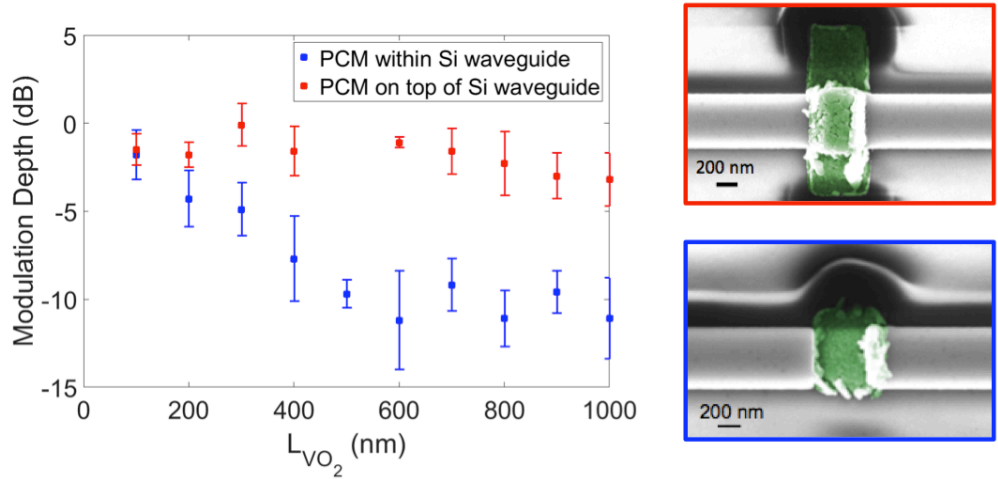

Fig. 6. Measured modulation depth for Si-VO$_2$ waveguide modulators with VO$_2$ on top of (red data points and upper right SEM image) and embedded within (blue data points and lower right SEM image) the Si waveguide.

While integration of PCMs within a Si waveguide has been proposed [25, 26], we believe this is the first experimental realization of such a structure. In [25], the authors select a thin slab of Ge$_2$Sb$_2$Te$_5$ (GST) as the PCM to be embedded within a Si waveguide. With a GST length of 2 μm, calculations suggest the Si-thin slab GST integrated waveguide would achieve an insertion loss of 0.6 dB and a modulation depth of 14 dB. Using GST as the PCM in our PCM embedded Si waveguide modulator geometry, for $L_{GST}$ = 200 nm, simulation results (GST optical properties taken from [27]) suggest similar metrics (i.e., insertion loss of 1.5 dB and modulation depth of 14.9 dB) could be achieved with one-tenth the modulator footprint of the geometry proposed in [25]. In addition to well-studied PCMs (i.e., VO$_2$ and GST), we believe the embedded trench approach is a promising platform to explore emerging, engineered PCMs with optimized properties, such as GSS4T1 [28], which by minimizing κ can further reduce insertion loss while maintaining a large modulation depth.

## 7. Conclusion

In summary, we demonstrate a compact, non-resonant, broadband hybrid Si-PCM integrated optical waveguide modulator by embedding a PCM within a Si waveguide for improved modal overlap with the PCM. With thermal activation, using VO$_2$ as the PCM, we report ~ 10 dB modulation at 1550 nm in experiment for a VO$_2$ length of only 500 nm. With fabrication improvements, calculations suggest modulation depth greater than 12 dB with less than 3 dB insertion loss can be achieved. In addition to VO$_2$, GST and other PCMs could be employed in the proposed integrated geometry to realize high-speed, small-footprint electro-optic and all-optical modulators for interchip and intrachip applications.

## Funding

This work was supported in part by the National Science Foundation (ECCS1509740).

## Acknowledgments

This research used resources (JEOL 9300FS 100kV, Oxford Plasmalab 100) at the Center for Nanophase Materials Sciences (CNMS), a DOE Office of Science User Facility operated by the Oak Ridge National Laboratory. This work also used resources (Raith eLine, Angstrom Engineering Amod A & B, Veeco Dektak 150) at the Vanderbilt Institute of Nanoscale Science and Engineering. The authors gratefully acknowledge technical assistance from Dr.